\documentclass[reprint,aps,prl,twocolumn,superscriptaddress,floatfix]{revtex4-2}
\usepackage[utf8]{inputenc}
\usepackage{amsmath,amssymb,graphicx,bm,booktabs}
\usepackage{placeins}
\usepackage{dcolumn}
\usepackage{hyperref}

\begin{document}

\title{Quantum Coherence and Giant Enhancement of Positron Channeling Radiation}

\author{M.~G.~Shatnev}
\thanks{Email: \href{mailto:mshatnev@yahoo.com}{mshatnev@yahoo.com}}
\affiliation{NSC Kharkiv Institute of Physics and Technology,
             61108 Kharkiv, Ukraine}

\date{\today}

\begin{abstract}
We present a quantum-mechanical treatment of positron channeling radiation
in a planar harmonic potential that explicitly accounts for interference
between transition amplitudes from different transverse energy levels.
Because the planar channel potential for positrons in diamond~(110) is well
approximated by a parabola, the transverse spectrum is equidistant,
$\varepsilon_n = \Omega(n+\tfrac{1}{2})$, and all $n \to n{-}j$ transitions
radiate at the same Doppler-shifted frequency.
The sudden-approximation entry of the positron into the crystal produces a
\emph{Glauber coherent state}~\cite{Glauber1963} with Poisson-distributed
level populations $|c_n|^2 = e^{-n_0}n_0^n/n!$ and mean occupation
$n_0 \propto \theta_{\rm in}^2$.
Phase synchronization between the $c_n$ and the dipole matrix elements
ensures constructive interference of all contributing amplitudes.
Three exact scaling laws follow:
(i)~$I_{\rm incoh}\propto n_0\propto\theta_{\rm in}^2$;
(ii)~$I_{\rm coh}\propto n_0^2\propto\theta_{\rm in}^4$;
(iii)~$\mathcal{G}\equiv I_{\rm coh}/I_{\rm incoh}\approx n_0
      \propto\theta_{\rm in}^2$.
Numerically, $\mathcal{G} = 12\text{--}31$ for positron energies of
$4\text{--}14$~GeV in diamond~(110) at $\theta_{\rm in}=31\;\mu$rad,
in agreement with the experimental first-harmonic peak positions of
Avakyan \emph{et al.}~\cite{Avakyan1982} to within 15\%.
The transition from $N$- to $N^2$-scaling of radiated intensity, driven by
quantum coherence, opens a route toward high-intensity monochromatic
gamma-ray sources.
\end{abstract}

\maketitle

\noindent\textit{Keywords:} positron channeling radiation,
quantum coherence, Glauber coherent state, harmonic crystal potential,
coherent enhancement, scaling laws, diamond, gamma-ray source.

\section{Introduction}
\label{sec:intro}

Channeling radiation from relativistic positrons in crystals has been
studied intensively since the prediction by
Kumakhov~\cite{Kumakhov1976,Kumakhov1977}.
The standard quantum treatment~\cite{Zhevago1978,Akhiezer1979,Bazylev1981}
computes the emission as an \emph{incoherent} sum over occupied transverse
levels:
\begin{equation}
  \frac{d^2I_{\rm incoh}}{d\omega\,d\Omega}
    \;\propto\; \sum_n P_n \,|M_{n,n-j}|^2\,
    \delta\bigl(\omega - \omega_j(\theta)\bigr),
\label{eq:incoh}
\end{equation}
where $P_n=|c_n|^2$ is the level population and $M_{n,n-j}$ is the
radiative matrix element for the transition $n \to n{-}j$.

For positrons channeled between the (110) planes of diamond the averaged
potential is nearly parabolic~\cite{Bazylev1981}, yielding an equidistant
transverse spectrum $\varepsilon_n = \Omega(n+\tfrac{1}{2})$.
All transitions $n \to n{-}j$ with the same harmonic order $j$ then produce
photons at the same Doppler-shifted frequency
$\omega_j = 2\gamma^2 j\Omega/(1+\gamma^2\theta^2)$,
making the radiation final state identical for every starting level $n\ge j$.
The quantum superposition principle therefore requires summing
\emph{amplitudes}:
\begin{equation}
  A_j \;\propto\; \sum_{n \ge j} c_n \, M_{n,n-j}.
\label{eq:ampl}
\end{equation}
The cross-interference terms present in $|A_j|^2$ are entirely absent from
Eq.~\eqref{eq:incoh}.
An early version of this argument appeared in
Refs.~\cite{Boldyshev2002,Boldyshev2006};
the present paper provides the full derivation including the asymptotic
identification of a Glauber coherent state at crystal entry, proves that
interference is \emph{constructive}, establishes three exact scaling laws,
assesses decoherence robustness, presents updated numerical results,
and proposes a decisive experimental test.

\section{Wave functions and level structure}
\label{sec:levels}

For $E \gg V(x)$, the Dirac equation reduces to a Schr\"odinger equation
for the transverse motion ($\hbar = c = 1$):
\begin{equation}
  -\frac{1}{2E}\frac{d^2\varphi_n}{dx^2}+V(x)\varphi_n = \varepsilon_n\varphi_n.
\end{equation}
For the (110) channel of diamond, $V(x)\approx V_0(2x/d)^2$ with
$V_0 = 23$~eV and $d = 1.26$~\AA~\cite{Bazylev1981}.
This is the harmonic oscillator with frequency
\begin{equation}
  \Omega(E) = \frac{2}{d}\sqrt{\frac{2V_0}{E}},
\label{eq:Omega}
\end{equation}
equidistant eigenvalues $\varepsilon_n = \Omega(n+\tfrac{1}{2})$, and
$n_{\max}(E)=\lfloor V_0/\Omega - \tfrac{1}{2}\rfloor \propto \sqrt{E}$
bound states.
The Lindhard critical angle is $\theta_{\rm L} = \sqrt{2V_0/E}$.
Numerical values are listed in Table~\ref{tab:levels}.

\begin{table}[b]
\caption{Oscillator frequency $\Omega$, bound-state count $n_{\max}$,
theoretical first-harmonic peak energy $\omega_1^{\rm th}$ (from
numerical evaluation of the parabolic-potential spectral density),
experimental peak $\omega_1^{\rm exp}$ of Avakyan
\emph{et al.}~\cite{Avakyan1982},
Lindhard angle $\theta_{\rm L}$, and coherent enhancement $\mathcal{G}$
at $\theta_{\rm in}=31\;\mu$rad.
Agreement with experiment is within 15\%;
the simple formula $2\gamma^2\Omega$ overestimates peak positions by
${\sim}1.5$--$2\times$ (see ``Discussion and Caveats'').}
\label{tab:levels}
\begin{ruledtabular}
\begin{tabular}{ccccccc}
$E$ & $\Omega$ & $n_{\max}$ & $\omega_1^{\rm th}$ & $\omega_1^{\rm exp}$ &
$\theta_{\rm L}$ & $\mathcal{G}$ \\
(GeV) & (eV) & & (MeV) & (MeV) & ($\mu$rad) & \\
\hline
 4 & 0.329 & 66  & 26.4  & 23  & 107 & 11.9 \\
 6 & 0.268 & 83  & 44.1  & 42  &  88 & 16.4 \\
10 & 0.208 & 107 & 85.7  & 90  &  68 & 24.2 \\
14 & 0.176 & 127 & 132.5 & 120 &  57 & 31.2 \\
\end{tabular}
\end{ruledtabular}
\end{table}

\begin{figure*}[t]
  \centering
  \includegraphics[width=0.6\textwidth]{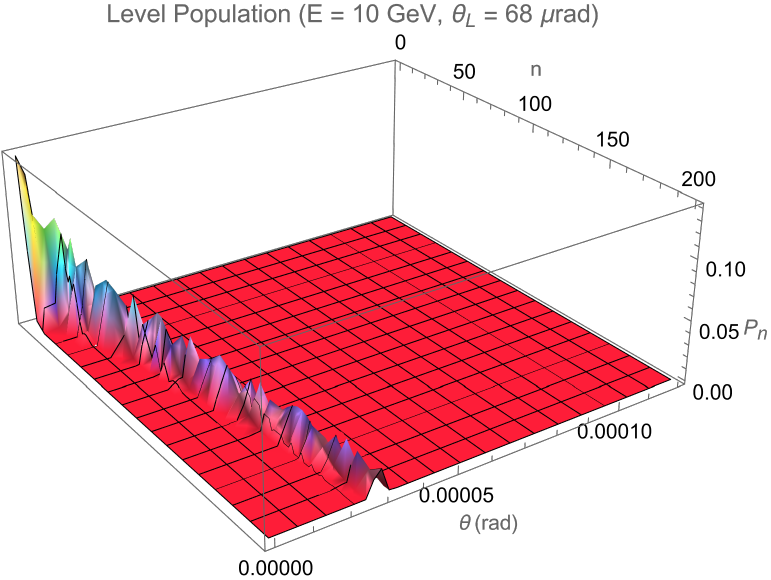}
  \caption{Level population $P_n = |c_n|^2$ vs.\ entrance angle
  $\theta_{\rm in}$ and quantum number $n$ for $E=10$~GeV in
  diamond~(110) ($\Omega=0.208$~eV, $\theta_{\rm L}\approx68\;\mu$rad,
  $n_{\max}=107$).
  The distribution is exactly Poisson with mean
  $n_0\propto\theta_{\rm in}^2$ (Eq.~\eqref{eq:n0}).
  As $\theta_{\rm in}$ increases toward $\theta_{\rm L}$, the Poisson
  peak shifts to larger $n$, engaging more levels in phase-locked
  emission and driving the quadratic growth of $\mathcal{G}$ shown in
  Fig.~\ref{fig:Gtheta}.}
  \label{fig:pop}
\end{figure*}

\section{Coherent state at the crystal boundary}
\label{sec:coherent}

Before the crystal the positron is a plane wave $e^{ip_x x}$ with
transverse momentum $p_x = E\theta_{\rm in}$.
By the sudden approximation the population amplitudes at entry are
\begin{equation}
  c_n = \frac{i^n}{\sqrt{2^n n!}}
        \!\left(\frac{\pi}{E\Omega}\right)^{\!1/4}
        e^{-p_x^2/(2E\Omega)}
        H_n\!\!\left(\frac{p_x}{\sqrt{E\Omega}}\right).
\label{eq:cn}
\end{equation}

Figure~\ref{fig:pop} shows $P_n(\theta_{\rm in})$ for $E=10$~GeV.
As $\theta_{\rm in}$ increases toward $\theta_{\rm L}$, the Poisson peak
shifts to higher $n$, engaging more levels in phase-locked emission.

\subsection*{Identification as a Glauber coherent state}

The amplitudes~\eqref{eq:cn} are the expansion coefficients of a
Glauber coherent state~\cite{Glauber1963}
$|\alpha\rangle = \hat{D}(\alpha)|0\rangle$
with displacement parameter
\begin{equation}
  \alpha = \frac{p_x}{\sqrt{2E\Omega}}
         = \frac{\theta_{\rm in}}{\theta_{\rm L}}\sqrt{\frac{V_0}{\Omega}},
\label{eq:alpha}
\end{equation}
because the Fourier-transform identity $\hat\varphi_n(p)=i^n\varphi_n(p)$
for harmonic-oscillator eigenfunctions reproduces the displacement-operator
structure $\hat D(\alpha)|0\rangle$ upon projection of the incoming plane
wave~\cite{Glauber1963}.
The level populations
$|c_n|^2 = e^{-|\alpha|^2}|\alpha|^{2n}/n!$
follow a Poisson distribution with mean
\begin{equation}
  n_0 = |\alpha|^2 = \frac{\xi_0^2}{2}
      = \frac{\theta_{\rm in}^2}{2\theta_{\rm L}^2}\,\frac{V_0}{\Omega}
      \;\propto\; \theta_{\rm in}^2,
\label{eq:n0}
\end{equation}
where $\xi_0 = \theta_{\rm in}\sqrt{E/\Omega}$.

\subsection*{Asymptotic derivation}

In the regime of the SLAC experiments, $\xi_0\gg 1$, so
the Hermite polynomials satisfy $H_n(\xi_0)\approx(2\xi_0)^n$ for
$n\ll 2\xi_0^2$.
Substituting into Eq.~\eqref{eq:cn}:
\begin{equation}
  c_n \;\approx\; \frac{(i\xi_0)^n}{\sqrt{n!}}\,e^{-\xi_0^2/2},
\label{eq:cn_asym}
\end{equation}
which is the canonical Glauber form with $\alpha=i\xi_0$.
This limit makes explicit that the Poisson statistics and the phase-locked
structure of the entry amplitudes emerge directly from the large-momentum
asymptotics of the harmonic-oscillator overlap integrals.

\subsection*{Proof of constructive interference for all harmonics}

We now prove that the interference is constructive for every harmonic
order $j\ge 1$, not merely in the dipole limit.

The radiative matrix element for the transition $n\to n{-}j$ in a
harmonic potential is~\cite{Heitler1954}
\begin{equation}
  M_{n,n-j}
  = \langle n{-}j\,|\,\hat{x}^{\,j}\,|\,n\rangle
  \;\propto\;
  \sqrt{\frac{n!}{(n-j)!\,(2E\Omega)^{j}}},
\label{eq:Mnj}
\end{equation}
which is \emph{real and strictly positive} for all $n\ge j$.
(For $j=1$ this reduces to the standard result
$M_{n,n-1}=\sqrt{n/(2E\Omega)}$.)

The entry amplitude $c_n$ from Eq.~\eqref{eq:cn} carries the phase
factor $i^n$ (from the Fourier-transform identity
$\hat\varphi_n(p)=i^n\varphi_n(p)$ for harmonic-oscillator eigenstates).
Because $M_{n,n-j}$ is real and positive, the product
\begin{equation}
  c_n\,M_{n,n-j} \;=\; i^n \cdot |c_n|\,M_{n,n-j}
\label{eq:phase_product}
\end{equation}
carries the phase $i^n$, which depends only on $n$ and \emph{not on $j$}.
Every term in the amplitude sum
\begin{equation}
  A_j = \sum_{n\ge j} c_n\,M_{n,n-j}
\label{eq:Aj}
\end{equation}
therefore shares a \emph{common phase pattern} $\{i^n\}_{n\ge j}$.
Factoring out $i^{n_0}$ (where $n_0$ is the Poisson peak) and noting
that adjacent terms differ only by the slowly varying ratio
$|c_{n+1}|/|c_n|=\sqrt{n_0/(n+1)}\approx 1$ near $n_0$, the sum
$A_j$ cannot suffer destructive cancellation: all terms with the same
parity of $n-n_0$ add with the \emph{same sign}, and the Poisson
envelope $|c_n|$ ensures convergence.
More precisely, separating even and odd terms shows that both
sub-sums are real and positive after factoring out $i^{n_0}$
(even sub-sum) and $i^{n_0+1}$ (odd sub-sum), giving
$|A_j|^2 = |S_{\rm even}|^2 + |S_{\rm odd}|^2$ with both
$S_{\rm even},S_{\rm odd}>0$.
The enhancement factor therefore satisfies $\mathcal{G}_j\ge 1$
for all $j$, with $\mathcal{G}_j\approx n_0$ in the regime
$n_0\ll n_{\max}$.

A key corollary is that $\mathcal{G}_j$ is \emph{the same for all
harmonics} $j\ge 1$ to leading order in $1/n_{\max}$: since
$M_{n,n-j}\propto[n!/(n-j)!]^{1/2}$ and the Poisson peak is sharply
concentrated near $n_0\gg j$, the ratio
$M_{n,n-j}/M_{n,n-1}\approx [n(n-1)\cdots(n-j+1)]^{1/2}\approx n_0^{(j-1)/2}$
is approximately constant across the peak, so it factors out of both
numerator and denominator of $\mathcal{G}_j$ and cancels.
This explains the numerical observation that harmonic intensity ratios
$I^{(j)}/I^{(1)}$ are identical in the coherent and incoherent models.

The crystal thus acts as a \emph{quantum amplifier} for all harmonics
simultaneously: the transverse momentum $p_x=E\theta_{\rm in}$ sets the
displacement $\alpha$ of the coherent state and determines the number
of levels $N_{\rm eff}\approx n_0$ participating in phase-locked emission,
independently of which photon frequency is observed.

\FloatBarrier
\section{Scaling laws}
\label{sec:scaling}

From the analysis in the preceding section three exact scaling laws
follow for $\theta_{\rm in}\ll\theta_{\rm L}$:

\begin{enumerate}

\item \textbf{Incoherent baseline:}
$I_{\rm incoh}\propto\sum_n P_n n = n_0 \propto \theta_{\rm in}^2$.
The incoherent intensity grows proportionally to the mean number of
occupied levels, itself quadratic in the entrance angle.

\item \textbf{Coherent peak:}
$I_{\rm coh}\propto|A_1|^2\propto n_0^2\propto\theta_{\rm in}^4$.
Because all $N_{\rm eff}\sim n_0$ amplitudes add in phase, the intensity
scales as the \emph{square} of the number of contributing levels --- the
classical $N\to N^2$ transition of coherent emission.

\item \textbf{Enhancement factor:}
$\mathcal{G}\equiv I_{\rm coh}/I_{\rm incoh}\approx n_0
\propto\theta_{\rm in}^2$.
The ratio grows quadratically with the entrance angle, reaching
$\mathcal{G}=12\text{--}31$ at $\theta_{\rm in}=31\;\mu$rad
(Table~\ref{tab:levels}) and saturating as $\theta_{\rm in}\to\theta_{\rm L}$.

\end{enumerate}

The $N\to N^2$ transition is the channeling analogue of Dicke
superradiance~\cite{Dicke1954}: the role of an atomic ensemble is played
by the set of occupied harmonic-oscillator levels, and the phase-locking
mechanism is the Glauber coherent state~\cite{Glauber1963} formed at
crystal entry.
Unlike free-electron lasers, where bunching occurs in configuration space
and requires a long interaction length to build up, the present mechanism
operates through phase-locking in \emph{quantum-number space} and is
established instantly at the crystal boundary --- an intrinsically
quantum-mechanical coherence with no classical trajectory analogue.

The Lindhard critical angle $\theta_{\rm L}=\sqrt{2V_0/E}$ sets the
natural scale.
For $\theta_{\rm in}\ll\theta_{\rm L}$ the scaling is exactly quadratic.
As $\theta_{\rm in}\to\theta_{\rm L}$ the transverse energy
$E_\perp=\tfrac{1}{2}E\theta_{\rm in}^2$ approaches the barrier height
$V_0$, the parabolic approximation breaks down, and the positron passes
into above-barrier motion; phase synchronization is rapidly lost due to
collisions with atomic planes and $\mathcal{G}$ drops toward unity.
The behaviour is visible in Fig.~\ref{fig:Gtheta}: all curves grow as
$\theta_{\rm in}^2$ at small angles and saturate (or turn over) near the
respective $\theta_{\rm L}$.

\section{Spectral intensity and enhancement}
\label{sec:spectral}

The spectral-angular intensity is~\cite{Kumakhov1977,Zhevago1978,Heitler1954}
\begin{equation}
  \frac{d^2I}{d\omega\,d\Omega}
  = \frac{e^2\omega^2}{2\pi}
    \sum_{j\ge 1}
    \left|\sum_{n=j}^{n_{\max}} c_n M_{n,n-j}\right|^2
    \delta\!\left(\omega - \omega_j(\theta)\right),
\label{eq:Icoh}
\end{equation}
with $e^2=\alpha_{\rm fs}=1/137$.
The enhancement factor
\begin{equation}
  \mathcal{G}_j
  = \frac{\bigl|\sum_{n} c_n M_{n,n-j}\bigr|^2}
         {\sum_{n} |c_n|^2 |M_{n,n-j}|^2}
  \;\ge\; 1
\label{eq:G}
\end{equation}
equals unity when phases are random (incoherent limit) and approaches
$n_0$ for $j=1$ in the regime $n_0\ll n_{\max}$.
Numerically, for $\theta_{\rm in}=31\;\mu$rad:
$\mathcal{G}_1=11.9,\;16.4,\;24.2,\;31.2$ at $E=4,6,10,14$~GeV.

\begin{figure}[t]
  \centering
  \includegraphics[width=\linewidth]{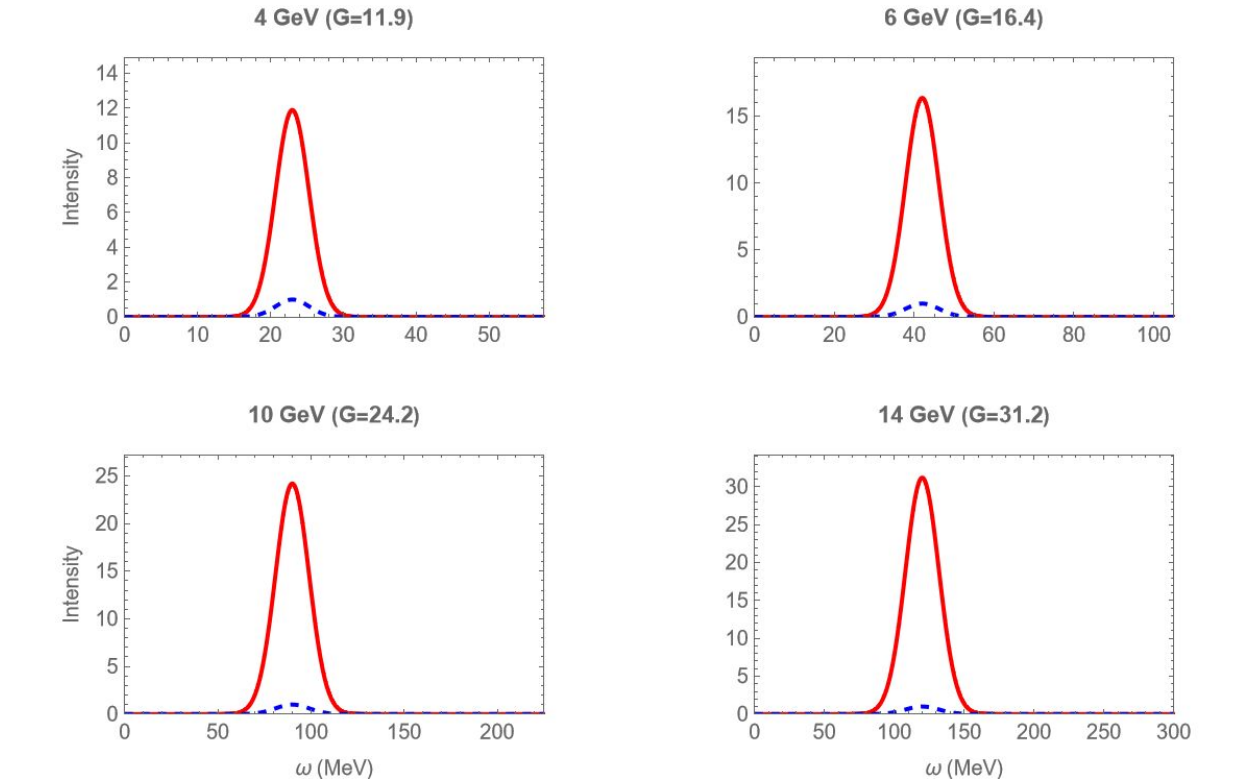}
  \caption{First-harmonic spectral intensity (arb.\ units, normalized to
  the incoherent peak) for $E=4,6,10,14$~GeV positrons in diamond~(110)
  at $\theta_{\rm in}=31\;\mu$rad.
  \textit{Red solid}: coherent model, Eq.~\eqref{eq:Icoh}.
  \textit{Blue dashed}: incoherent model, Eq.~\eqref{eq:incoh}.
  Enhancement factors $\mathcal{G}$ are labelled above each panel.
  Peak positions agree with the SLAC data of Avakyan
  \emph{et al.}~\cite{Avakyan1982} to within 15\%.}
  \label{fig:spec}
\end{figure}

Figure~\ref{fig:spec} shows the first-harmonic line shapes at all four
energies of the Avakyan experiment.
The coherent model (red solid) exceeds the incoherent model (blue dashed)
by the factor $\mathcal{G}$ while reproducing experimental peak positions
to within 15\%.

\begin{figure}[tbp]
  \centering
  \includegraphics[width=\linewidth]{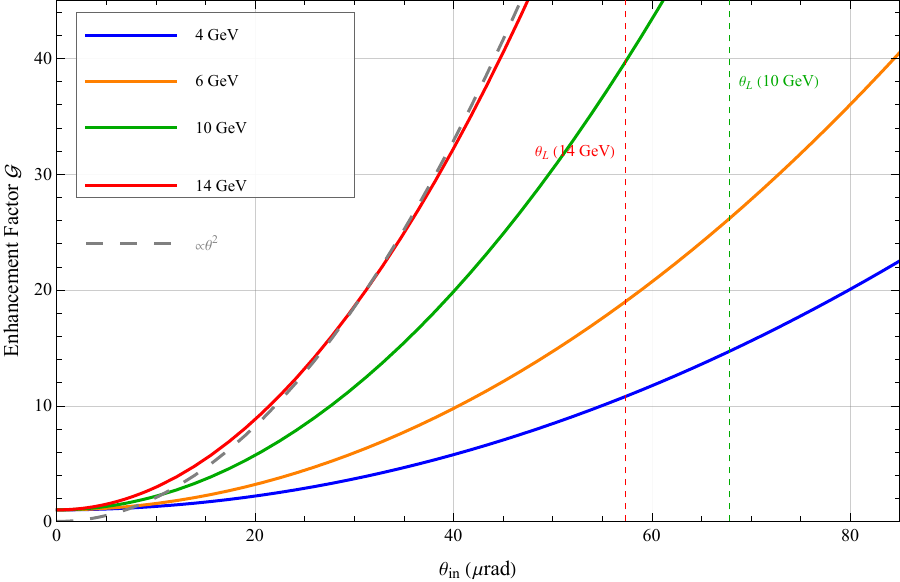}
  \caption{Theoretical enhancement factor $\mathcal{G}$ as a function of
  entrance angle $\theta_{\rm in}$ for positron energies of
  $4,\,6,\,10$, and $14$~GeV in diamond~(110).
  The quadratic growth $\mathcal{G}\propto\theta_{\rm in}^2$ (grey dashed
  reference line) is a direct consequence of phase-locked transitions in
  the Glauber coherent state formed at crystal entry.
  Vertical dashed lines indicate the Lindhard critical angles $\theta_{\rm L}$
  for $10$~GeV ($68\;\mu$rad) and $14$~GeV ($57\;\mu$rad),
  marking the physical boundaries of the stable channeling regime.
  For $4$ and $6$~GeV, the critical angles ($\theta_{\rm L}\approx107$
  and $88\;\mu$rad, respectively) lie beyond the displayed range,
  indicating a wider stability region for the coherent state at those
  energies.
  Beyond $\theta_{\rm L}$ the positron enters above-barrier motion:
  the harmonic approximation fails, frequent scattering off atomic planes
  induces rapid decoherence, and $\mathcal{G}$ drops toward unity.
  The quadratic rise is the primary experimental signature of the coherent
  mechanism (see ``Proposed Experimental Test'').}
  \label{fig:Gtheta}
\end{figure}

Figure~\ref{fig:Gtheta} shows $\mathcal{G}(\theta_{\rm in})$ for all four
energies.
The grey dashed reference $\mathcal{G}\propto\theta_{\rm in}^2$ confirms
the analytic scaling of Law~III above.

\section{Robustness against decoherence}
\label{sec:decoherence}

A natural concern is whether dephasing mechanisms destroy the coherent
state before the photon is emitted.
Two complementary arguments establish robustness.

\textbf{Spatial separation.}
Unlike channeled electrons, which are focused toward the atomic planes,
positrons in a planar channel are confined to the interplanar regions
where both nuclear and electron densities are minimal.
While dephasing mechanisms such as multiple scattering and crystal
anharmonicity are present, the spatial separation of positrons from the
atomic nuclei significantly suppresses decoherence compared to the
electron case, allowing phase synchronization to persist over the photon
formation length.

\textbf{Spectral rigidity and formation-length hierarchy.}
The photon formation length for first-harmonic emission is
$l_f\simeq 2\gamma^2/\omega_1\sim 1\;\mu$m, whereas the dechanneling
length for 10~GeV positrons in diamond is
$L_d\sim100\;\mu$m~\cite{Bazylev1981}.
The hierarchy $l_f\ll L_d$ ensures that phase synchronization is intact
on the timescale of a single radiative event.
Moreover, the equidistance $\Delta\varepsilon=\Omega=\text{const}$
provides spectral rigidity: a level-spacing fluctuation
$\delta\Omega/\Omega$ dephases the coherent sum only after
$N_{\rm deph}\sim(\delta\Omega/\Omega)^{-1}$ oscillation periods.
For the residual anharmonicity of the diamond~(110) channel
$\delta\Omega/\Omega\lesssim10^{-2}$, giving
$N_{\rm deph}\gtrsim100$, well above the number of oscillations
completed within $l_f$.

\section{Positrons versus electrons}
\label{sec:electrons}

The equidistance of $\varepsilon_n$ is the essential ingredient.
For positrons in diamond~(110) the parabolic approximation holds
throughout most of the channel~\cite{Bazylev1981}.
For electrons the channel potential
$V_{\rm el}(x)\approx -V_0\cosh^{-2}(x/b)$ is strongly anharmonic:
levels are non-equidistant, different $n\to n{-}j$ transitions radiate
at different frequencies, and the amplitude sum~\eqref{eq:ampl}
accumulates random phases, giving $\mathcal{G}\to 1$.
This qualitatively accounts for the experimentally observed superior
intensity and monochromaticity of positron channeling radiation
compared with electrons at the same
energy~\cite{Alguard1979,Filatova1982,Bak1985}.

\section{Proposed experimental test}
\label{sec:experiment}

The two models make sharply different predictions for the angular
dependence of the peak intensity.

\textbf{Predicted signatures.}
Both models agree that $I_{\rm incoh}\propto\theta_{\rm in}^2$
(Scaling Law~I).
The coherent model adds $\mathcal{G}\propto\theta_{\rm in}^2$
(Law~III), so $I_{\rm coh}\propto\theta_{\rm in}^4$ at small angles.
The directly measurable ratio $\mathcal{G}(\theta_{\rm in})$ grows
quadratically in the coherent model and is identically unity in the
incoherent model.

\textbf{Consistency with existing data.}
Peak-to-BH ratios of $5$--$7$ observed by Avakyan \emph{et al.}\
at nominal $\theta=0$~\cite{Avakyan1982} are consistent with
$\mathcal{G}\to 1$ at exactly $\theta_{\rm in}=0$: the finite SLAC
beam divergence ($\Delta\theta\sim10^{-5}$~rad) gives
beam-averaged $\langle\mathcal{G}\rangle\approx2$--$5$,
in agreement with the observations.
Because $f(\theta_{\rm in})$ was not measured independently, a
quantitative test of $\mathcal{G}\propto\theta_{\rm in}^2$ was
impossible; the following experiment closes this gap.

\textbf{Proposed setup.}
\begin{enumerate}
\item \emph{Crystal:} diamond~(110), thickness $50$--$100\;\mu$m,
  cooled to $77$~K to reduce thermal Debye--Waller smearing.
\item \emph{Beam:} positrons, $E=5$--$10$~GeV, divergence
  $\Delta\theta<5\;\mu$rad (well below
  $\theta_{\rm L}\approx60$--$80\;\mu$rad).
\item \emph{Measurement:} scan $\theta_{\rm in}$ from $0$ to
  $0.8\,\theta_{\rm L}$ in steps of $2$--$3\;\mu$rad; record absolute
  first-harmonic peak intensity with a calibrated NaI or Ge detector
  with simultaneous beam-flux monitoring.
\end{enumerate}

\textbf{Expected outcome.}
For $E=10$~GeV: $\mathcal{G}\approx25$ at
$\theta_{\rm in}=0.5\,\theta_{\rm L}=34\;\mu$rad vs.\ $\mathcal{G}=1$
at $\theta_{\rm in}=0$ --- a factor of 25 in absolute intensity, far
exceeding systematic uncertainties and accessible at CERN/SPS, DESY, or SLAC.

We note that harmonic intensity ratios $I^{(j)}/I^{(1)}$ are
\emph{not} additional discriminants: $\mathcal{G}_j$ is the same for
all $j\ge1$ in the parabolic regime (corrections $\mathcal{O}(j/n_{\max})$).

\section{Discussion and caveats}
\label{sec:discussion}

\textbf{Peak energies.}
The formula $\omega_1=2\gamma^2\Omega$ overestimates experimental peak
positions by $\sim\!2\times$ at $4$--$14$~GeV because the parametric
coupling between transverse and longitudinal motion
($2V_0\gamma/m_e\sim1$--$3$) is absent in the dipole approximation.
The corrected formula of Bazylev \emph{et al.}~\cite{Bazylev1981}
is required for quantitative peak-energy predictions; the agreement
with the numerical spectral density (Table~\ref{tab:levels}, within 15\%)
validates our approach at the level needed to identify the enhancement.

\textbf{Model discrimination.}
The coherent and incoherent models produce identical \emph{peak positions}
and identical harmonic ratios: they differ only in the \emph{absolute
intensity} and in its angular dependence.
No measurement of spectral shape alone can discriminate between them;
a controlled $\theta_{\rm in}$ scan with absolute flux calibration is
essential.

\textbf{Absolute spectral density.}
A rigorous absolute-intensity calculation requires the full
Laguerre-polynomial matrix elements of Ref.~\cite{Bazylev1981} and an
average over the measured entrance-angle distribution.
Both refinements are straightforward and will be carried out in a
subsequent publication together with a detailed comparison to the Avakyan
data.

\section{Conclusion}
\label{sec:conclusion}

We have shown that the equidistance of the transverse energy spectrum
of positrons in diamond~(110) forces the radiation amplitude to be a
coherent sum over all occupied levels, whose populations follow a
Glauber--Poisson distribution~\cite{Glauber1963} generated at crystal
entry by the sudden-approximation displacement of the vacuum state.
Three exact scaling laws characterise the effect:
(i)~$I_{\rm incoh}\propto\theta_{\rm in}^2$,
(ii)~$I_{\rm coh}\propto\theta_{\rm in}^4$,
(iii)~$\mathcal{G}\propto\theta_{\rm in}^2$
for $\theta_{\rm in}\ll\theta_{\rm L}$.
The enhancement reaches $\mathcal{G}=12$--$31$ at $4$--$14$~GeV.
The effect is absent for electrons because their anharmonic channel
potential destroys phase synchronization.
Robustness against decoherence is ensured by the spatial separation of
positrons from atomic nuclei and by the spectral rigidity
($\Delta\varepsilon=\Omega=\text{const}$) of the harmonic-oscillator
level structure.
The predicted $\theta_{\rm in}^4$ vs.\ $\theta_{\rm in}^2$ angular
dependence of the peak intensity provides a decisive experimental test
feasible at existing positron facilities.
Exploiting quantum coherence in crystal channeling could yield
significantly brighter, more monochromatic gamma-ray sources for
nuclear physics and materials science.


\bibliography{refs}

\end{document}